\documentclass[prb,twocolumn,floatfix,showpacs]{revtex4}
\usepackage{psfrag,epsfig} 
\usepackage{graphicx}
\usepackage{amssymb}
\usepackage{amsmath}
\usepackage{feynmf}
\begin{document}
\title{Renormalization group study of Luttinger liquids with boundaries} 
\author{S.~Grap} \affiliation{Institut f\"ur Theoretische Physik A and JARA--Fundamentals of Future Information Technology, RWTH Aachen University, 
52056 Aachen, Germany}  
\author{V.~Meden} \affiliation{Institut f\"ur Theoretische Physik A and JARA--Fundamentals of Future Information Technology, RWTH Aachen University, 
52056 Aachen, Germany}

\date{October 21, 2009}

\begin{abstract}
We use Wilsons weak coupling ``momentum'' shell renormalization group method to show that 
two-particle interaction terms commonly neglected in bosonization of one-dimensional correlated 
electron systems with open boundaries are indeed irrelevant in the renormalization group 
sense. Our study provides a more solid ground for many investigations of Luttinger liquids 
with open boundaries. 
\end{abstract}
\maketitle     

{\it Introduction---}The electron-electron interaction has strong effects on the low-energy physics
of (quasi) one-dimensional (1d) metals. Such systems cannot be described within 
Landaus Fermi liquid (FL) theory but rather form a different ``universality'' class: 
the Luttinger liquids (LLs).\cite{Schoenhammer05} Already decades ago the importance of local 
single-particle inhomogeneities in LLs allowing for electron backscattering with 
momentum transfer $2k_F$ (were $k_F$ is the Fermi momentum) was 
emphasized.\cite{LutherPeschel,Mattis} Their relevance 
became even more apparent after it was shown that on low energy scales a LL 
(with repulsive two-particle 
interaction) with a single local impurity in many respects behaves as if the chain was cut in two at 
the position of the impurity with open boundary conditions at the endpoints.\cite{KaneFisher,Furusaki0}       
This led to a number of studies on the physics of LLs with open boundaries\cite{FG,EMJ,WVP,VWG,MMX} mainly 
using the method of bosonization.\cite{Schoenhammer05,Herbert}

The Tomonaga-Luttinger model\cite{Schoenhammer05} is a translationally invariant (no inhomogeneities) 
continuum model within the LL ``universality'' class for which all correlation functions can 
be computed exactly by bosonization. In the context of bulk LL physics it plays a role similar to 
the noninteracting electron gas for FL physics.\cite{Schoenhammer05} In this model the $2k_F$ 
two-particle scattering process of spin-up and spin-down electrons is neglected. 
All the other
low-energy scattering process can be written as quadratic forms in the bosonic densities 
of the right- (around $k_F$) and left-moving (around $-k_F$) fermions. Furthermore, after 
linearization of the single-particle dispersion, also the kinetic energy is quadratic in the bosonic 
densities\cite{Schoenhammer05,Herbert} and the remaining Hamiltonian is that of noninteracting bosons. 
Neglecting the above $2 k_F$ scattering process leads to 
an additional conservation law, namely the ``local'' (in momentum space) spin around the two Fermi 
points $\pm k_F$, which is central to the exact solution of the Tomonaga-Luttinger model. 
In the literature one finds 
two arguments why this so-called $g_{1,\perp}$-term can be neglected. Either one follows the original 
idea of Tomonaga\cite{Tomonaga} and assumes a two-particle interaction which is sufficiently smooth and
long-ragend
in real-space (suppressed screening) such that its Fourier component at $2 k_F$ vanishes 
or, more generally, one adopts the 
so called g-ology renormalization group (RG) approach of  S\'olyom.\cite{solyom} 
In this it is shown that  in an important part of the parameter space $g_{1,\perp}$ is 
irrelevant in the RG sense. The solution $g_{1,\perp}=0$ is a stable fixed point of the RG 
flow.\cite{solyom} To (qualitatively) 
understand the low-energy physics the $ g_{1,\perp}$-term can thus be neglected. It only affects the 
numerical values of the other fixed-point couplings.\cite{foot1} 

For open boundary conditions the single-particle 
quantum number $k$ in which the noninteracting problem is diagonal (see below) no longer 
corresponds to the momentum. Therefore the two-particle scattering terms appearing in this natural 
basis have a form different from those of the  translationally invariant
case. 
Also for LLs with open boundary conditions the Hamiltonian contains two-particle scattering terms  
which cannot be written as quadratic forms in bosonic densities, when considering a general, 
spin-conserving two-particle interaction. 
It was shown\cite{MMX} that these terms vanish if 
Tomonagas rather specific assumption of an interaction which is smooth in real-space is made. 
Then open-boundary bosonization can be used to compute all correlation functions of the 
open-boundary analog of the Tomonaga-Luttinger model.\cite{FG,EMJ,WVP,VWG,MMX,Dirk} Surprisingly, 
an RG analysis for more general two-particle interactions similar to S\'olyoms approach was so 
far not discussed in the standard literature on open-boundary bosonization. In the present Brief 
Report we close this gap. Using Wilsons ``momentum'' shell RG in weak 
coupling\cite{foot2,Shankar} we show that all two-particle scattering terms naturally arising in a 
low-energy analysis of a 1d system with general two-particle interaction and open boundaries 
which cannot be written as quadratic forms in the bosonic densities are RG irrelevant (open boundary 
g-ology analysis). We demonstrate that the RG flow equations are the same as the  
ones of the translationally invariant g-ology model. In analogy to this the RG irrelevant coupling 
constants only affect the fixed-point couplings of the terms quadratic in the bosons and can thus be 
neglected for a qualitative understanding of the low-energy physics of LLs with open boundaries. Our 
analysis puts many of the studies of LLs with open boundaries on a more solid ground. 

{\it Model---}To be specific we consider the 1d electron gas on a line between $x=0$ and $x=L$ 
with a general, sufficiently regular two-particle interaction 
\begin{eqnarray*}
\hat V  & = &  \frac{1}{2} \sum_{s,s'} \int^L_0 \!\!\!\! dx \int^L_0 \!\!\!\!  dx'
\psi_{s}^{\dag}(x) \psi_{s'}^{\dag}(x') \\
&& \times V_{s,s'}(x-x') \psi_{s'}^{}(x') \psi_{s}^{}(x) 
\end{eqnarray*}     
where $s,s'$ denote the spin. The noninteracting one-particle eigenstates are given by 
standing waves (open boundaries) $  \varphi_n (x) = \sqrt{2/L} \sin{ (k_n x)} $, with 
quantum numbers $k_n = n\pi/L$,  $n \in {\mathbb N}$. We emphasize that for open boundaries only 
right-moving electrons with a positive Fermi velocity $v_F=k_F/m_e$ appear, where $m_e$ denotes the 
electron mass. The field operator is given by 
$\psi_{s}^{(\dag)}(x) = \sum_{n=1}^{\infty} \varphi_n (x) a^{(\dagger)}_{n,s} $ 
with the creation and annihilation operators $ a^{(\dag)}_{n,s}$ of electrons in the noninteracting 
eigenstates. We now closely follow
Ref.\ \onlinecite{MMX} to rewrite the interacting part of the Hamiltonian as 
\begin{eqnarray*}
\hat V  = \frac{1}{2} \sum_{s,s'} \sum_{n,m,n',m'} v^{s,s'}_{n,m,n',m'}  
a^{\dag}_{n,s}  a^{\dag}_{m,s'} a_{m',s'}  a_{n',s} 
\end{eqnarray*} 
with the matrix elements 
\begin{eqnarray}
\mbox{} \hspace{-.3cm} && v_{n,m,n',m'}^{s,s'} = 
\left[ F_{s,s'} \left(k_n-k_{n'},k_m-k_{m'}\right) \right.
\nonumber \\ 
\mbox{} \hspace{-.3cm} && - F_{s,s'}\left(k_n-k_{n'},k_m + k_{m'}\right)  
- F_{s,s'} \left(k_n+k_{n'},k_m - k_{m'}\right) \nonumber \\
\mbox{} \hspace{-.3cm} && \left. + F_{s,s'}\left(k_n + k_{n'},k_m + k_{m'}\right) \right]/L \; ,
\label{Mele}
\end{eqnarray}
where 
\begin{eqnarray*}
F_{s,s'}(q,q')  = 
\frac{1}{L} 
\int^L_0 \!\!\!\! dx  \int^L_0 \!\!\!\! dx' \cos{( qx)} V_{s,s'}(x-x')   \cos{(q'x')} \; .
\end{eqnarray*}
\vspace{-.8cm}
\begin{eqnarray}
\label{Fdef}
\end{eqnarray}
The kinetic part of the Hamiltonian $\hat H = \hat T + \hat V$ is given by 
\begin{eqnarray*}
\hat T = \sum_s \sum_n \epsilon_{k_n} a^{\dag}_{n,s} a_{n,s}
\end{eqnarray*}
with the single-particle dispersion $\epsilon_k = k^2/(2m_e)$.

{\it Perturbative Wilson RG---}To analyze the Hamiltonian $\hat H$ using a Wilson RG we consider 
the action $S(\bar \Psi,\Psi)$ appearing in the imaginary time functional integral representation 
of the grand canonical partition function
\begin{eqnarray*}
&& S(\bar \Psi,\Psi) =  S_0 + S_I \\
&& =  - \sum_s  \sum_{n} \sum_j  \bar \Psi_s(k_n,\omega_j) [{\mathcal G}_0(k_n,\omega_j)]^{-1} 
\Psi_s(k_n,\omega_j) \\
&& + \beta^{-1} \sum_{s,s'} \sum_{n,m,n',m'} \sum_{i,j,i',j'}  \delta_{i+j,i'+j'} 
v^{s,s'}_{n,m,n',m'} \\ && \times  \bar \Psi_s(k_n,\omega_i)  \bar \Psi_{s'}(k_m,\omega_j)  \Psi_{s'}(k_{m'},\omega_{j'}) 
 \Psi_{s}(k_{n'},\omega_{i'}) \; ,
\end{eqnarray*}
with Gra{\ss}mann variables $\Psi$, $\bar \Psi$, Matsubara frequencies $\omega_j$, the 
noninteracting propagator 
${\mathcal G}_0(k,\omega)= [i \omega - \epsilon_k + \mu]^{-1}$ ($\mu$ is the chemical potential), 
and the inverse temperature $\beta$. Here we will only be interested in the behavior at zero temperature 
($\beta \to \infty$) with $\mu=\epsilon_F=k_F^2/(2m_e)$. In this case we obtain $\epsilon_k - \mu = 
v_F (k-k_F) + (k-k_F)^2/(2m_e)$.

We now follow the standard steps of the perturbative Wilson $k$-shell RG:\cite{Shankar} (i) separating
the Gra{\ss}mann fields  $\Psi= \Psi^< + \Psi^>$ into fast modes $ \Psi^>$ (relative to the Fermi point) 
with $\Lambda/b\leq |k-k_F| \leq \Lambda$ and slow ones $\Psi^<$, 
(ii) integrating out the fast modes $\Psi^>$ perturbatively, (iii) rescaling the quantum numbers $\tilde k = k-k_F$ 
($\tilde k'=b \tilde k$)  and the Matsubara frequencies ($\omega'=b \omega$)\cite{foot3} and (iv) 
rescaling the Gra{\ss}mann fields  $\Psi'(\tilde k'+k_F,\omega') = \xi^{-1} 
\Psi^<(\tilde k'/b+k_F,\omega'/b)$. 

We first consider the noninteracting case. Then the action can naturally be written as 
$S_0 = S_0^<(\bar\Psi^<,\Psi^<) + S_0^>(\bar\Psi^>,\Psi^>)$ and the second step of the RG procedure 
only leads to a constant in the partition function. Therefore the rescaling is the only relevant step. 
Choosing $\xi = b^{1/2}$ the $i \omega - v_F (k-k_F)$ part of the noninteracting action remains 
invariant, while the quadratic term of the dispersion vanishes as $1/b$ for large $b$. It is 
RG irrelevant for the low-energy physics and from now on we work with ${\mathcal G}_0(k,\omega)= 
[i \omega - v_F (k-k_F)]^{-1}$. The standard linearization can thus be justified by the RG. Note that 
to avoid proliferation of symbols we do not introduce a new one for the free propagator of the
model with linearized dispersion. 

Next we study the flow of the quartic part of the action. As usual in a weak coupling RG to compute 
$\left< e^{-S_I} \right>_{S_0^>}$, where the index refers to taking the expectation value with respect 
to the fast mode part of the noninteracting action, we use a cumulant expansion
\begin{eqnarray*}
\ln \left< e^{-S_I} \right>_{S_0^>} = -\left<  S_I \right>_{S_0^>} + \frac{1}{2} \left(  
\left<  S_I^2 \right>_{S_0^>} - \left<  S_I \right>_{S_0^>}^2 \right) + \ldots \; . 
\end{eqnarray*}
\vspace{-.8cm}
\begin{eqnarray}
\label{cumu}
\end{eqnarray}
On tree level (first term of the above expansion) and for low energy scales ($b \to \infty$) 
the RG steps lead similarly to the translational invariant case to a two-particle interaction 
which is purely local in real space 
$V_{s,s'}(x-x') \to V_{s,s'}(0) \delta(x-x')$. All correction terms 
are subleading.\cite{Shankar} Then the integrals in Eq.\ (\ref{Fdef}) can be performed leading to 
\begin{eqnarray*}
F_{s,s'}(q,q')  = V_{s,s'}(0) \left( \delta_{q,q'} + \delta_{q,-q'} \right)/2 \; .
\end{eqnarray*}
This has to be contrasted to the case with periodic boundary conditions in which only the second 
Kronecker delta appears and the interaction matrix element only has the contribution coming from
$F_{s,s'}(k_n-k_{n'},k_m-k_{m'})$ with $k$ being the momentum.\cite{MMX}
Since for open boundaries all the incoming and outgoing quantum numbers must be close to the 
single Fermi point only the Kronecker deltas $ \delta_{n-{n^\prime},m-{m^\prime}},
\delta_{n-{n^\prime},-m+{m^\prime}}$, and $\delta_{n+{n^\prime},m+{m^\prime}}$ in the expression 
Eq.\ (\ref{Mele}) for the interaction matrix element contribute. As in the translationally 
invariant g-ology model we now generalize the interaction and assume  different coupling constants for 
each of the remaining low-energy scattering processes and each of the relative spin orientations 
of the two particles. Doing so one can expect that the following low-energy RG analysis is applicable not 
only to the electron gas model but also of relevance for a larger class of models, including 
lattice ones.  For the interaction matrix element Eq.\ (\ref{Mele}) this leads to 
\begin{eqnarray*}
 2 L v_{n,m,n',m'}^{s,s'} & = & 
(\sideset{_{||}}{_1}{\mathop{g}} \delta_{s,s^{\prime
}}+\sideset{_{\perp}}{_1}{\mathop{g}} \delta_{s,-s^{\prime}}) 
\delta_{n+{n^\prime},m+{m^\prime}} \\
&& + (\sideset{_{||}}{_2}{\mathop{g}} \delta_{s,s^{\prime
}}+\sideset{_{\perp}}{_2}{\mathop{g}} \delta_{s,-s^{\prime}}) 
\delta_{n-{n^\prime},m-{m^\prime}} \\
&& + \left (  \sideset{_{||}}{_4}{\mathop{g}} \delta_{s,s^{\prime}}  + 
\sideset{_{\perp}}{_4}{\mathop{g}} \delta_{s,-s^{\prime}} \right) 
\delta_{n-{n^\prime},-m+{m^\prime}}  \; .
\end{eqnarray*} 
The initial values of the generalized coupling constants are functions of the 
microscopic parameters of the underlying model. The relation between these two parameter sets can 
(in principle) be determined by an additional RG step or simpler by perturbation 
theory. For the coupling constants we use the same 
notation\cite{solyom}  $g_1$, $g_2$ and $g_4$ as in the translationally 
invariant g-ology model,\cite{foot4} 
while we still make clear that we are dealing 
with a different situation by putting the spin orientation index to the left of the $g$'s.\cite{relspin} 
Two reasons for this become clear already at this stage of the discussion. In analogy to 
the standard g-ology the $\sideset{_{\perp}}{_1}{\mathop{g}}$-term is the one which 
cannot be written as a quadratic form in the bosonic 
density operators of the right-moving electrons (for a detailed discussion on this, see 
Ref.\ \onlinecite{MMX}). Also in strict analogy the $\sideset{_{||}}{_2}{\mathop{g}}$- and 
$\sideset{_{||}}{_1}{\mathop{g}}$-terms describe the same scattering processes (as can be seen 
by reordering Gra{\ss}mann fields and renaming indices in the expression for $S$) 
and could be combined in a single coupling constant $\sideset{_{||}}{_2}{\mathop{g}} - 
\sideset{_{||}}{_1}{\mathop{g}}$ (see also the RG flow equations below). 

\begin{figure}[t]

 \includegraphics[width=6.8cm]{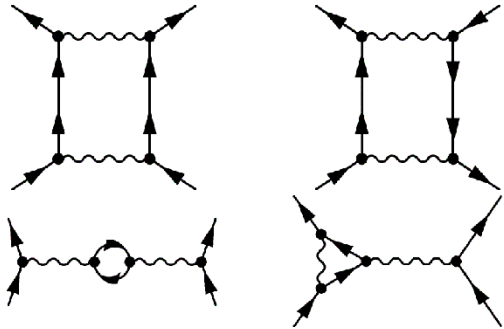}

\caption{Topology of the second order diagrams contributing to the RG flow.}
\label{fig1}
\end{figure}

With these interaction matrix elements one now has to compute all (second order) connected diagrams 
of the two-particle vertex related to the second term of Eq.\ (\ref{cumu}) with all the
external quantum numbers 
$k$ set to $k_F$ and the Matsubara frequencies set to zero. Writing down the corresponding analytical 
expressions 
it becomes immediately clear that only the one-particle irreducible diagrams give nonvanishing 
contributions. Furthermore, in complete analogy to the g-ology RG of the translationally invariant 
model to leading order the $g_4$-terms do not contribute to the RG flow of the coupling constants 
(they do not lead to $\ln(b)$-terms) and do not flow themself. They only lead to a renormalization of 
the Fermi velocity, which is  a higher order effect if the RG flow of the couplings is considered.
We thus neglect the $g_4$-term. In a straight 
forward but tedious calculation we evaluated all the remaining second order diagrams 
with the topologies as shown in Fig.\ \ref{fig1}. The wiggled line stands for any of 
the four coupling constants $\sideset{_{||}}{_1}{\mathop{{\gamma}}}$, 
$\sideset{_{\perp}}{_1}{\mathop{{\gamma}}}$, $\sideset{_{||}}{_2}{\mathop{{\gamma}}}$, and 
$\sideset{_{\perp}}{_2}{\mathop{{\gamma}}}$. We identified if a certain diagram gives a 
finite RG flow of the vertex and if so to which of the four scattering channels it 
contributes. To perform the remaining $k$-sum we took the limit $L \to \infty$ (resulting in a 
semi-infinite chain). Setting $b=e^s$ and 
resorting to infinitesimal RG steps this leads to the RG flow equations
\begin{eqnarray}
\frac{d}{ds} \sideset{_{||}}{_1}{\mathop{{\gamma}}}&=& -\left({\sideset{_{||}}{_1}{\mathop{{\gamma}}}}^2 +
{\sideset{_{\perp}}{_1}{\mathop{{\gamma}}}}^2 \right) \; , \nonumber \\
\frac{d}{ds}  \sideset{_{||}}{_2}{\mathop{{\gamma}}}&=&- {\sideset{_{||}}{_1}{\mathop{{\gamma}}}}^2  \; ,\nonumber\\
\frac{d}{ds} \sideset{_{\perp}}{_1}{\mathop{{\gamma}}}&=&-2 \sideset{_{\perp}}{_1}{\mathop{{\gamma}}} 
\left( \sideset{_{\perp}}{_2}{\mathop{{\gamma}}}  - \sideset{_{||}}{_2}{\mathop{{\gamma}}}
+ \sideset{_{||}}{_1}{\mathop{{\gamma}}}\right) \; ,\nonumber\\
\frac{d}{ds} \sideset{_{\perp}}{_2}{\mathop{{\gamma}}}&=&- {\sideset{_{||}}{_1}{\mathop{{\gamma}}}}^2 \; , 
\label{PBCflow1}
\end{eqnarray}
with 
\begin{eqnarray*}
\sideset{_{\nu}}{_l}{\mathop{{\gamma}}} = \frac{ \sideset{_{\nu}}{_l}{\mathop{{g}}} }{\pi v_F}
\end{eqnarray*}
for $\nu = ||, \perp$ and $l=1,2$. The coupling constants $\gamma$ are now understood to be 
functions of the infrared cutoff parameter $s$.

\begin{figure}[tb]
   \includegraphics[height=5.8cm]{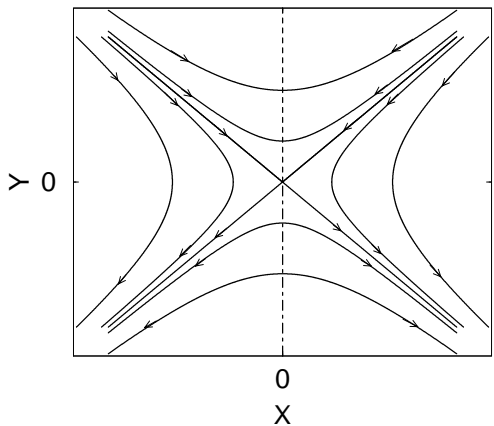}
   \caption{Sketch of the trajectories of the weak coupling RG flow in the 
  $\sideset{_{\perp}}{_1}{\mathop{{\gamma}}}$-$\gamma_{1,2}$-plane 
  ($X=\sideset{_{\perp}}{_1}{\mathop{{\gamma}}}$, $Y=\gamma_{1,2}$).  The dashed line indicates 
a line of stable fixed points. The fixed points on the dashed-dotted line are unstable.}
   \label{fig2}
\end{figure}

Remarkably, these equations are exactly the lowest-order RG flow equations 
for the coupling constants (divided by $2 \pi v_F$ instead of $\pi v_F$) of the 
translationally invariant g-ology model.\cite{solyom}  This provides the complete justification 
for the decision to use the same notation. For the solution of the RG equations and an analysis 
of the fixed points we can proceed as it is well documented in the literature.\cite{solyom} 
We first introduce the coupling constant
\begin{eqnarray*}
\gamma_{1,2} = \sideset{_{||}}{_1}{\mathop{{\gamma}}} - \sideset{_{||}}{_2}{\mathop{{\gamma}}}
+ \sideset{_{\perp}}{_2}{\mathop{{\gamma}}} \; ,
\end{eqnarray*}      
which leads to the RG flow equations 
\begin{eqnarray*}
\frac{d}{ds} \gamma_{1,2} & = & -2 {\sideset{_{\perp}}{_1}{\mathop{{\gamma}}}}^2  \; ,\\
\frac{d}{ds} \sideset{_{\perp}}{_1}{\mathop{{\gamma}}}&=&-2 \sideset{_{\perp}}{_1}{\mathop{{\gamma}}} 
\gamma_{1,2} \; .
\end{eqnarray*}
From this it becomes apparent that $\gamma_{1,2}^2 -   
{\sideset{_{\perp}}{_1}{\mathop{{\gamma}}}}^2$ is invariant under the RG flow which implies that the 
RG trajectories form hyperbolas in 
the $\sideset{_{\perp}}{_1}{\mathop{{\gamma}}}$-$\gamma_{1,2}$-plane. The equations can easily be 
solved and trajectories are sketched in Fig.\ \ref{fig2}. If initially (for the initial cutoff) 
all coupling constants are small and
$\gamma_{1,2}^i \geq 
\left| {\sideset{_{\perp}}{_1}{\mathop{{\gamma}}}}^i \right|$ holds they all stay small (the use 
of perturbative RG is justified) and the flow is towards a line of stable 
fixed  points with 
\begin{eqnarray}
&& { \sideset{_{\perp}}{_1}{\mathop{{\gamma}}}}^\ast =0 \; ,\;\;\; \gamma_{1,2}^\ast = 
\left([\gamma_{1,2}^i]^2 - {[{ \sideset{_{\perp}}{_1}{\mathop{{\gamma}}}}^i]}^2 \right)^{1/2} \; , \nonumber \\ 
&& { \sideset{_{\perp}}{_2}{\mathop{{\gamma}}}}^\ast =  {\sideset{_{\perp}}{_2}{\mathop{{\gamma}}}}^i + \frac{1}{2} 
\left( \gamma_{1,2}^\ast  - \gamma_{1,2}^i \right) 
\label{FPwerte} 
\end{eqnarray}
indicated by the 
dashed line in Fig.\ \ref{fig2}. 
Under the above restriction on the initial coupling constants the term which cannot be written 
as a quadratic form in the bosonic density is RG irrelevant. The initial value 
$ {\sideset{_{\perp}}{_1}{\mathop{{\gamma}}}}^i$ of the
$ \sideset{_{\perp}}{_1}{\mathop{{\gamma}}}$-term  only affects the fixed-point values of the 
other couplings.\cite{solyom} The use of the open boundary analog of the Tomonaga-Luttinger model 
is then justified for studies on the low-energy physics and standard 
open-boundary bosonization\cite{FG,EMJ,WVP,VWG,MMX} can be used. This allows to compute all 
correlation functions.\cite{FG,EMJ,WVP,VWG,MMX} In the resulting 
expressions the coupling constants 
$\sideset{_{\perp}}{_2}{\mathop{{\gamma}}}$, $\sideset{_{||}}{_2}{\mathop{{\gamma}}}$
and $\sideset{_{||}}{_1}{\mathop{{\gamma}}}$
must be replaced by the fixed-point values Eq.\ (\ref{FPwerte}).

For a model in which the coupling constants on the initial scale of the weak coupling RG 
do not depend on the relative orientation of the spins of the two scattering electrons (like 
the electron gas model) one finds $\gamma_{1,2} = \sideset{_{||}}{_1}{\mathop{{\gamma}}}$ 
and $\sideset{_{\perp}}{_1}{\mathop{{\gamma}}} = \sideset{_{||}}{_1}{\mathop{{\gamma}}}$. 
Therefore the above condition on the initial couplings  for reaching a stable fixed 
point reduces to the simple requirement that the interaction must be repulsive. In this case 
the trajectory in Fig.\ \ref{fig2} flows to the origin.

{\it Summary---}We have shown that under quite general assumptions on the two-particle 
interaction the terms which are usually ignored in bosonization studies on LLs with open boundaries 
are indeed RG irrelevant. These terms which cannot be written as quadratic forms in the bosonic 
densities only affect the fixed-point values of the other couplings but do not modify 
the low-energy physics. For this analysis we have used a perturbative Wilson RG scheme. Although the 
scattering terms for open boundaries are of different nature than in the translationally invariant 
standard g-ology model we have shown that the RG equations have exactly the same form. Our result
puts a large number of bosonization studies on LL with open boundaries on a more solid ground. The 
present discussion is limited to the weak coupling regime. Numerical studies show 
that a similar low-energy physics can also be found at larger (repulsive) couplings.\cite{MMX}      
We finally note that exactly the same results can be obtained using the functional renormalization
group\cite{HM} instead of the Wilson RG.\cite{dipl}

{\it Acknowledgments---}We thank K.\ Sch\"onhammer and S.\ Andergassen for very 
valuable discussions, C. Karrasch for careful reading of the manuscript, and the Deutsche 
Forschungsgemeinschaft for support (FOR723).


\begin{thebibliography}{*}



\bibitem{Schoenhammer05}
For a recent review on LL physics see K.~Sch\"onhammer in 
{\it Strong Interactions in Low  Dimensions,} Eds.: D.~Baeriswyl 
and L.~Degiorgi, Kluwer Academic Publishers, Dordrecht (2005).
\bibitem{LutherPeschel} A.\ Luther and I.\ Peschel, 
Phys.\ Rev.\ B {\bf 9}, 2911 (1974).
\bibitem{Mattis} D.C.\ Mattis, J.\ Math.\ Phys.\ {\bf 15}, 609 (1974). 
\bibitem{KaneFisher} 
C.L.\ Kane and M.P.A.\ Fisher, Phys.~Rev.~Lett. {\bf 68}, 1220 (1992); 
Phys.\ Rev.\ B {\bf 46}, 15233 (1992).
\bibitem{Furusaki0} A.~Furusaki and N.~Nagaosa, Phys.~Rev.~B {\bf 47}, 
4631 (1993).
\bibitem{FG} M.\ Fabrizio and A.\ Gogolin, Phys.\ Rev.\ B {\bf  51}, 
17827 (1995).
\bibitem{EMJ} S.\ Eggert, H.\ Johannesson, and A.\ Mattsson,
Phys.\ Rev.\ Lett.\  {\bf 76}, 1505 (1996).
\bibitem{WVP} Y.\ Wang, J.\ Voit, and F.-C.\ Pu, Phys.\ Rev.\ B 
{\bf 54}, 8491 (1996). 
\bibitem{VWG} J.\ Voit, Y.\ Wang, and M.\ Grioni, Phys.\ Rev.\ B {\bf 61},
  7930 (2000).  
\bibitem{MMX}  V.~Meden, W.~Metzner, U.~Schollw\"ock, 
 O.~Schneider, T.~Stauber, and K.~Sch\"on\-hammer,
 Eur. Phys. J. B {\bf 16}, 631 (2000).
\bibitem{Dirk} For the discussion of correlation functions in the presence 
of boundaries and a bulk energy gap see D.~Schuricht, F.H.L.~Essler, 
A.~Jaefari, and E.~Fradkin, Phys.\ Rev.\ Lett.\ {\bf 101}, 086403 (2008).
\bibitem{Herbert} For an introduction to bosonization see J.\ von Delft and 
H.\ Schoeller, Ann.\ Phys.\ (Leipzig) {\bf 7}, 225 (1998). 
\bibitem{Tomonaga} S.\ Tomonaga, Prog.\ Theor.\ Phys.\ {\bf 5}, 544 (1950).
\bibitem{solyom} J.\ S\'olyom, Adv.\ Phys.\ {\bf 28}, 201 (1979). 
\bibitem{foot1} We here and in the following speak of ``couplings'' (and not of coupling functions)
as the $k$-dependence  turns out to be RG irrelevant (besides the classification 
concerning the momentum transfer; $2 k_F$ or $0$ for the translational invariant case).\cite{solyom}
\bibitem{foot2} For open boundaries the quantum number $k$ does not correspond to the 
momentum. In the following we thus speak of the ``Wilson $k$-shell RG''. 
\bibitem{Shankar} For a review on the application of Wilson RG to interacting Fermi systems 
see R. Shankar, Rev.\ Mod.\ Phys.\ {\bf 66}, 129 (1994). 
\bibitem{foot3} To obtain a fixed-point scenario rescaling of the Matsubara frequencies is 
mandatory.
\bibitem{foot4} The umklapp scattering term $g_3$ appearing in the translationally invariant 
g-ology model if an underlying half filled lattice model is assumed\cite{solyom} is irrelevant 
for our considerations.  
\bibitem{relspin} We note that using the index $\perp$ for the couplings in which two electrons with 
antiparallel spin scatter is to some extend misleading. As this is the notation widely used in the literature 
we here nevertheless follow this convention. 
\bibitem{HM} M.\ Salmhofer and C.\ Honerkamp, Prog.\ Theor.\ Phys.\ {\bf 105}, 1 (2001).
\bibitem{dipl} S.\ Grap, Diploma thesis, RWTH Aachen University (2009).
\end{thebibliography}
\end{document}